\documentclass[conference]{IEEEtran}
\IEEEoverridecommandlockouts
\usepackage{cite}
\usepackage{amsmath,amssymb,amsfonts}
\usepackage{algorithmic}
\usepackage{graphicx}
\usepackage{textcomp}
\usepackage{xcolor}
\def\BibTeX{{\rm B\kern-.05em{\sc i\kern-.025em b}\kern-.08em
    T\kern-.1667em\lower.7ex\hbox{E}\kern-.125emX}}

\begin{document}

\title{Hybrid Paradigm-based Brain-Computer Interface for Robotic Arm Control\\

\thanks{This work was partly supported by Institute of Information \& Communications Technology Planning \& Evaluation (IITP) grant funded by the Korea government (MSIT) (No. 2017-0-00432, Development of Non-Invasive Integrated BCI SW Platform to Control Home Appliances and External Devices by User’s Thought via AR/VR Interface; No. 2017-0-00451, Development of BCI based Brain and Cognitive Computing Technology for Recognizing User’s Intentions using Deep Learning; No. 2019-0-00079, Artificial Intelligence Graduate School Program(Korea University))}
}

\author{\IEEEauthorblockN{Byeong-Hoo Lee}
\IEEEauthorblockA{\textit{Dept. Brain and Cognitive Engineering} \\
\textit{Korea University}\\
Seoul, Republic of Korea \\
bh\_lee@korea.ac.kr}
\and
\IEEEauthorblockN{Jeong-Hyun Cho}
\IEEEauthorblockA{\textit{Dept. Brain and Cognitive Engineering} \\
\textit{Korea University}\\
Seoul, Republic of Korea \\
jh\_cho@korea.ac.kr}
\and
\IEEEauthorblockN{Byoung-Hee Kwon}
\IEEEauthorblockA{\textit{Dept. Brain and Cognitive Engineering} \\
\textit{Korea University}\\
Seoul, Republic of Korea \\
bh\_kwon@korea.ac.kr}
}

\maketitle

\begin{abstract}
Brain-computer interface (BCI) uses brain signals to communicate with external devices without actual control. Particularly, BCI is one of the interfaces for controlling the robotic arm.  In this study, we propose a knowledge distillation-based framework to manipulate robotic arm through hybrid paradigm induced EEG signals for practical use. The teacher model is designed to decode input data hierarchically and transfer knowledge to student model. To this end, soft labels and distillation loss functions are applied to the student model training. According to experimental results, student model achieved the best performance among the singular architecture-based methods. It is confirmed that using hierarchical models and knowledge distillation, the performance of a simple architecture can be improved. Since it is uncertain what knowledge is transferred, it is important to clarify this part in future studies.
\end{abstract}

\begin{small}
\end{small}

\IEEEpeerreviewmaketitle

\section{Introduction}
Motion imagery (MI) \cite{MI, MI3,jeong2020multimodal} is a widely used paradigm in brain-computer interfaces (BCIs), which induces meaningful brain signals through imagination that operates without actual movements. Since MI is related to movements, some studies on controlling external devices using MI-based BCI have been conducted \cite{wheelchair, roboticarm2, MI2}. Speech imagery (SI) is related to imagining speaking without actually speaking. Therefore SI has the advantage of being able to intuitively generate brain signals \cite{Nguyen, tian2016mental, Torres-Garcia}. Considering the characteristics of the two paradigms, using these paradigms together would be advantageous for a practical BCI system. In this study, we designed hybrid paradigm-based BCI that uses MI and SI to control robotic arm. Specifically, MI is in charge of the robotic arm movements, and SI is in charge of communication. 

Numerous studies have been conducted to improve SI and MI classification performance. For MI classification, Ang \textit{et al.} \cite{FBCSP} designed filter bank that contains several spatial patterns to select features by searching optimal filters for classification called FBCSP. Schirrmeister \textit{et al.} \cite{deepconvnet} designed convolutional neural network (CNN) which was inspired by pipeline of FBCSP. The experimental results showed that shallow architecture is advantageous for MI classification. Lawhern \textit{et al.} \cite{eegnet} applied depth-wise convolution and separable convolution that can avoid overparameterization. Fahimi \textit{et al.} \cite{Fahimi} proposed a generative adversarial network-based framework to generate additional EEG signals for performance improvement. For speech imagery classification, DaSalla $et \ al.$\cite{DaSalla} introduced common spatial pattern (CSP) for SI classification. Torres-Garcia $et\ al.$ \cite{Torres-Garcia} designed fuzzy inference to find Pareto front for selecting optimal channels. Nguyen $et\ al.$ \cite{Nguyen} showed that Riemannian manifold is efficient to classify long and short words. Their experiment showed complexity of word is the key factor of SI classification. 

In this paper, we proposed a hybrid paradigm-based BCI to efficiently control robotic arm. Furthermore, we applied knowledge distillation for compressing the model parameters that reduces computational complexity of the system. Teacher model is composed of three convolutional neural network (CNN) architectures that hierarchically classifies EEG signals. Student model is a singular CNN architecture. We assume that hierarchical classification has advantages in hybrid paradigm BCI, and through knowledge distillation, even a singular structure shows satisfactory classification performance. For evaluation, two datasets were used; SI and MI datasets of BCI competition 2020\footnote{$http://brain.korea.ac.kr/bci2021$}. The results of the experiment demonstrated that hybrid paradigm-based BCI is efficient for practical use through the knowledge distillation.

\begin{figure*}[!t]
  \centerline{\includegraphics[scale =0.9]{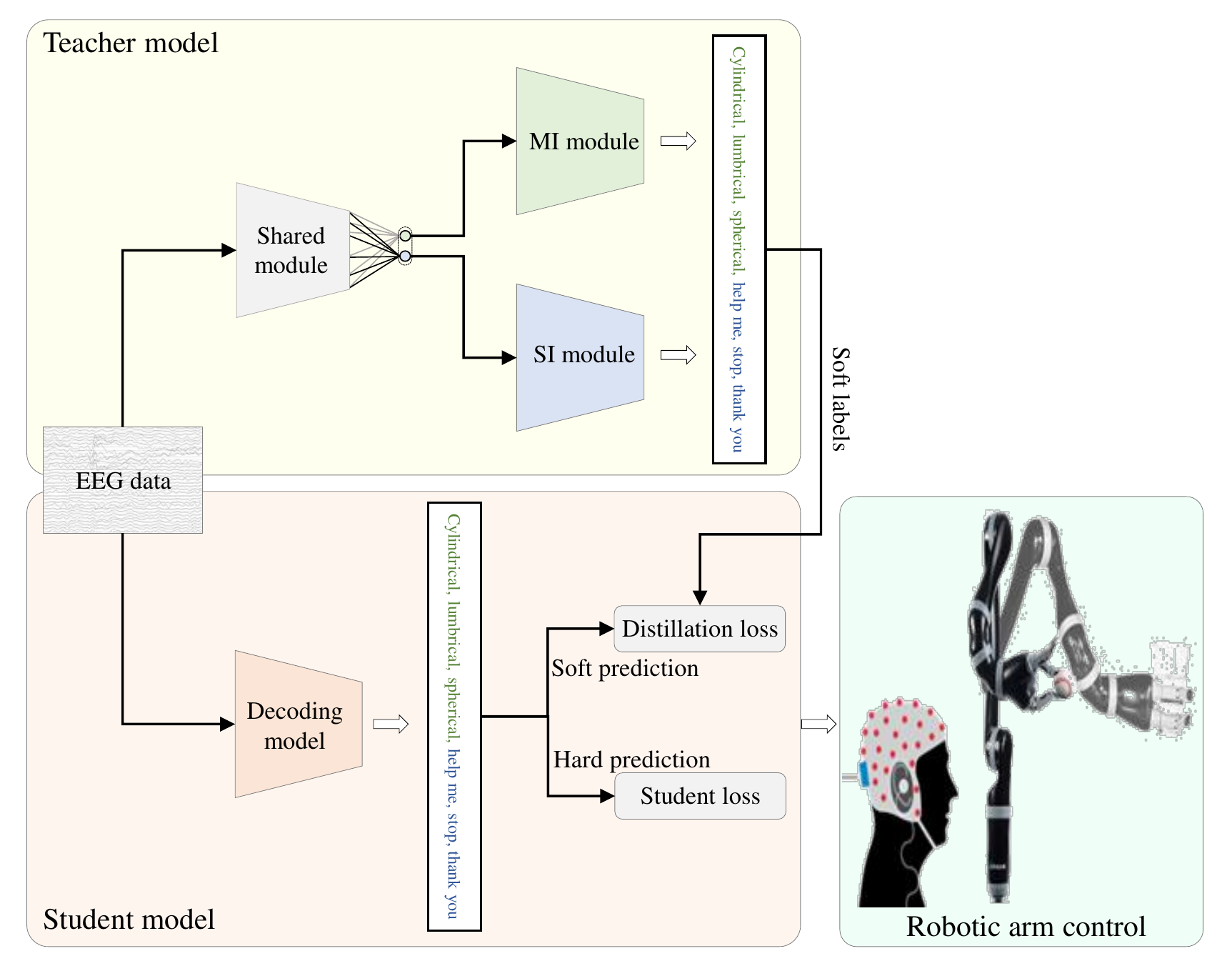}}
  \caption{Framework of the proposed method. The teacher model is learned using half the samples on the dataset. The student model is transferred knowledge from the teacher model through soft label and distillation loss. At this time, the teacher model is also trained using whole training set. In robotic arm control system, only student model operates as a decoding model.}
\end{figure*}

\begin{figure}[!t]
  \centerline{\includegraphics[scale =0.7]{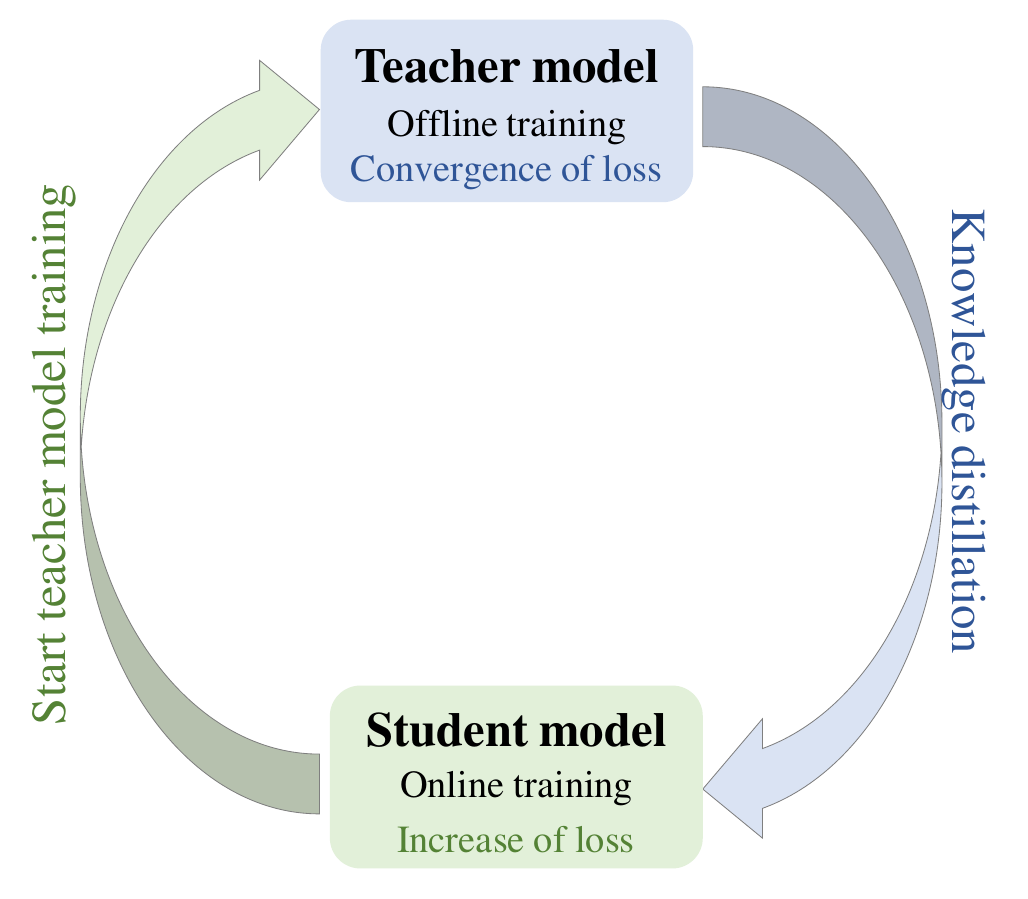}}
  \caption{Pipeline of the knowledge distillation. During student model training, teacher model transfer knowledge to student model until the loss converge. }
\end{figure}

\begin{table}[t!]
{\normalsize
\caption{Design choices of the proposed model. Details of student model is described in sub section $B. Student Model$}
\renewcommand{\arraystretch}{1.25}
\resizebox{\columnwidth}{!}{%
\begin{tabular}{llll} \hline
Parameter                     & Shared module        & MI module        & SI module       \\ \hline
Input       & Raw EEG             & Feature input           & Feature input           \\ 
                              & (1, 1, 24, 1000)     & (1, 36, 1, 288)          & (1, 36, 1, 288)          \\\hline
Hidden layer & Conv2D: 36, 72          & Conv2D: 72, 144, 288 & Conv2D: 72, 144, 288 \\
                              & AvgPool: (1,3)      & AvgPool: (1,3)           & AvgPool: (1,3)           \\
                              & Stride: (1,3)       & Stride: (1,3)            & Stride: (1,3)            \\ \hline
Activation   & ELU                 & ELU                      & ELU                      \\
                              & Last layer: Softmax & Last layer: Softmax      & Last layer: Softmax      \\ \hline
Optimizer                     & AdamW                & AdamW                     & AdamW                     \\ \hline
\end{tabular}}}
\end{table}

\section{Methods}
We propose a singular CNN structure-based robotic arm control system that decodes MI and SI-induced EEG signals. Basically, the EEG signals have no underlying ground truth of EEG signals and intricate and non-stationary signals characteristics that make it difficult to classify the MI and SI induced EEG signals. Moreover, since MI and SI induce EEG signals in different brain regions, decoding MI and SI-induced EEG signals is difficult using only singular architecture. Therefore, we applied the knowledge distillation technique by setting up a teacher model that classifies signals hierarchically and a student model that has a singular CNN architecture. Overall flow chart of the proposed method is described in Fig.1.

\subsection{Teacher Model}
Teacher model consists of three CNN architectures inspired by \cite{hier}; shared module, MI module, and SI module. The shared module consisted of two convolution layers to classify paradigm. The first layer performs temporal and spatial convolution by reducing channel dimension into single. The second layer performs convolution and categorizes paradigm defining it as MI or SI. According to prediction, MI or SI module utilizes the features from the second layer as input to conduct classifications for each paradigm. 

MI and SI modules exploit features to improve classification accuracy, specializing in SI and MI classification. Both SI and MI modules are composed of three convolution-pooling layers. During the model training two modules received features from the shared module regardless of the shared module's prediction. Through this, one module learns corresponding paradigm signals while the other module learns the wrong cases at the same time. Therefore, the modules are specialized in handling each paradigm EEG signals. Detailed design choices are described in Table I.

To train teacher model, we designed loss function that is composed of three loss terms \cite{lee2020classification} defined as follows:

\begin{equation}
loss(L_{sh}, L_M, L_S) = L_{sh} + p_M L_M + p_S L_S 
\end{equation}
where ${p_M}$ and ${p_S}$ are predictions of MI and SI for shared module, respectively. ${L_{sh}}$, ${L_M}$ and ${L_S}$ are cross-entropy loss \cite{entropy} of the shared module, MI module and SI module, respectively, which are defined as follow:

\begin{equation}
    \begin{split}
    L_{sh} = -\sum_{c=1}^{2}y_{sh}\log{\hat{y}_{sh}} \\
    L_M = -\sum_{c=1}^{M}y_{m}\log{\hat{y}_{m}} \\
    L_S = -\sum_{c=1}^{N}y_{s}\log{\hat{y}_{s}}
    \end{split}
\end{equation}
where ${y_{sh}}$, ${y_m}$, and ${y_s}$ are label of shared module, MI module, and SI module, respectively. On the other hand, ${\hat{y}_{sh}}$, ${\hat{y}_{m}}$, and ${\hat{y}_{s}}$ are prediction of the shared module, MI module, and SI module, respectively. $c$ denotes class, and the shared module conducts binary classification because it is a paradigm classifier.

\subsection{Student Model}
Three convolution-pooling layers composed of a student model. Basically, it has the same structure as the shared module, and we added a layer with a 144 size convolution filter. It was designed to reduce the interference time in robotic arm control system. The purpose of the student model is to distill the knowledge of the teacher model and utilize it for training to classify hybrid paradigm induced EEG signals even it is a simple structure.

\subsection{Knowledge Distillation}
Since the teacher model is a hierarchical structure, the computational cost is high. Therefore, for daily life BCI-based robotic arm control, a decoding model need to show robust performance while the computational cost is small. As mentioned above, we designed a student model with a singular architecture that is guided by the teacher model through knowledge distillation to achieve robust classification performance. To this end, we combined cross-entropy loss with distillation loss \cite{hinton2015distilling} which is defined as

\begin{equation}
L = \sum_{(x,y) \in D} L_{KD}(S(x,\theta_{S}, \tau), T(x, \theta_T , \tau)) + \lambda L_{CE} (\hat{y_S} ,y)
\end{equation}
where S, T, $x$, and $y$ denote student and teacher model, input, and true label. $\theta$, $\tau$, and $\hat{y_S}$ denote model parameter, temperature, and student model output. $T$ makes the output of the lower input larger and the output of the larger input smaller to maximize the benefits of using soft labels. Pipeline of the knowledge distillation is described in Fig.2.

\section{Results and Discussions}
We used BCI competition 2020 dataset composed of MI and SI dataset. MI dataset is composed of dataset that contains 150 trials and 3 classes (cylindrical, lumbrical, and spherical grasp). On the other hand, 300 trials and 5 classes consist of SI dataset. It contains 60 training trials and 10 trials of validation set and test set, respectively. Hence, MI dataset contains 150 trials including all datasets, we randomly selected 80 trials from the MI dataset for this study to avoid data imbalance problem. As the evaluation was conducted in leave-one-subject-out manner. Twenty four channels were selected because MI and SI datasets share 24 channels (Fp1 Fp2 F7 F3 Fz F4 T8 CP5 CP1 CP2 CP6 O2 AF7 AF3 AF4 AF8 C1 C2 C6 TP7 PO3 POz PO4 PO8). Evaluation was conducted with 0.001 learning rate, 5 iteration patience, 32 batch size, and 200 training epoch. AdamW \cite{AdamW} optimizer with weight decay (0.01) was applied for the experiment. Experimental environment was conducted as a subject independent manner on an Intel 3.60 Core i7 9700 K CPU with 32 GB of RAM, NVIDIA TITAN V GPU, CUDA/Cudnn, and Python version 3.9 with PyTorch version 1.9. 

\begin{table}[!t]
{\normalsize
\caption{Comparison of evaluation results. Reported accuracy was calculated leave-one-subject-out manner. The upper subscripts $^1$ and $^2$ denote deep and shallow ConvNet \cite{deepconvnet}, respectively. RF denotes random forest. The highest numbers are bold.}
\renewcommand{\arraystretch}{1}
\begin{tabular}{ccc} \hline
\multicolumn{1}{l}{} & \textbf{Paradigm}     & \textbf{Class}       \\
\textbf{Model}       & \textbf{Accuracy (std)} & \textbf{Accuracy (std)} \\\hline
CSP+RF \cite{qi2012random}       & 68.18 (3.52)            & 23.61 (6.89)            \\
CSP+SVM \cite{FBCSP}             & 67.74 (6.89)            & 22.98 (6.12)            \\
CSP+LDA \cite{FBCSP}             & 70.05 (2.64)            & 25.41 (9.10)            \\
FBCSP \cite{FBCSP}               & 74.57 (6.65)            & 45.47 (5.27)            \\
EEGNet \cite{eegnet}              & 84.64 (4.12)            & 54.56 (4.30)            \\
ConvNet$^1$ \cite{deepconvnet}     & 82.57 (3.33)            & 52.52 (4.02)            \\
ConvNet$^2$ \cite{deepconvnet}     & \textbf{86.98 (6.26)}            & 55.46 (4.72)            \\
Shared module     & 83.61 (3.81)            & -           \\
Teacher Model     & -           & \textbf{61.27 (4.03)}           \\
Student Model     & -           & 57.36 (5.29)           \\\hline
\end{tabular}}
\end{table}

CSP is a method of maximizing the variance of classes based on covariance matrices and minimizing the variance of other classes. Relatively small numbers of data lead to overfitting and performance degradation. Random forest, support vector machines, and linear discriminant analysis were selected as classifiers for CSP. Table II shows that CSP-based methods show lower accuracy (around 70\%) than CNN architectures (around 81.95\%) in paradigm classification. For class classification, their average accuracy was around 29.47\% which is 30\% lower than CNN-based methods (56.23\%). The dataset consists of two paradigm-induced EEG signals, and these dataset properties are expected to be a hindrance to the mechanism of CSP. Additionally, the small number of training samples can be one of the reasons of performance degradation. 

Four CNN-based methods \cite{deepconvnet, eegnet} including the proposed method were used. EEGNet \cite{eegnet} and shallow ConvNet \cite{deepconvnet} showed similar performance in both paradigm and class classification. Although shallow ConvNet was designed to extract band power features, which are known as MI-specific features, it showed the best accuracy in paradigm classification. Compared to paradigm classification, CNN-based methods dropped around 30\% in class classification. The teacher model was designed to decode hybrid paradigm induced EEG signals through hierarchical classification and achieved the highest performance (61.27\%) although shared module showed only 83.61\%. However, student model which was trained through knowledge distillation achieved second highest accuracy. It recorded the best performance among the singular CNN architectures. Through the results, it confirmed that the hierarchical architecture improves the performance of a singular architecture through knowledge distillation. Results are described in Table II.

However, it was not revealed how the student model was influenced by the knowledge of the teacher model through knowledge distillation. Furthermore, further experiments are needed on whether simply using soft labels can efficiently use the performance of student models. Nevertheless, considering that student models show higher performance than single-structure-based methods, knowledge distillation can be one way to improve the performance of hybrid paradigm-based BCIs.

\section{Conclusion and Future Works}
\label{sec:print}
In this paper, we proposed a knowledge distillation based-framework that was designed to decode hybrid paradigm induced EEG signals for robotic arm control. Teacher model was designed to hierarchically classify input data. Student model was trained under the guidance of teacher models through soft labels and distillation loss function. According to the results, knowledge distillation improves student model performance even it is a singular architecture. However, additional experiments are needed on how the student model was influenced by the teacher model through the distillation of knowledge. Thus, our future work is to ensure that the hierarchical model transfers efficient knowledge to a singular architecture through the distillation of knowledge.

\section{Acknowledgement}
The author would like to thank K. Yin and H.-B. Shin for help with the discussion of the data analysis.\\
\bibliographystyle{IEEEtran}
\bibliography{Reference}

\end{document}